\documentclass{article}
\usepackage{amsmath,graphicx,hyperref}
\usepackage{booktabs}
\usepackage{xcolor}
\usepackage{makecell}
\usepackage{amsmath}
\usepackage{amssymb}
\usepackage{multirow}
\usepackage{enumitem}
\usepackage[preprint]{spconf}

\title{KSDiff: Keyframe-Augmented Speech-Aware Dual-Path Diffusion for Facial Animation}

\name{Tianle Lyu$^{\dagger}$, Junchuan Zhao$^{\dagger}$, Ye Wang$^{\star}$
\thanks{
$^{\dagger}$ Equal contribution. \\  
\hspace*{1.5em} $^{\star}$ Corresponding author. Email: wangye@comp.nus.edu.sg.\\
\hspace*{1.5em} Emails: tianle\_lyu@u.nus.edu, junchuan@u.nus.edu.}
}

\address{School of Computing, National University of Singapore}

\copyrightnotice{%
  \begin{minipage}{\textwidth}
  \centering
  \footnotesize
    \textcopyright \ 20XX IEEE. Personal use of this material is permitted. Permission from IEEE must be obtained for all other uses, in any current or future media, including reprinting/republishing this material for advertising or promotional purposes, creating new collective works, for resale or redistribution to servers or lists, or reuse of any copyrighted component of this work in other works. 
  \end{minipage}
}

\begin{document}
\ninept
\maketitle

\begin{abstract}
Audio-driven facial animation has made significant progress in multimedia applications, with diffusion models showing strong potential for talking-face synthesis. However, most existing works treat speech features as a monolithic representation and fail to capture their fine-grained roles in driving different facial motions, while also overlooking the importance of modeling keyframes with intense dynamics. To address these limitations, we propose KSDiff, a Keyframe-Augmented Speech-Aware Dual-Path Diffusion framework. Specifically, the raw audio and transcript are processed by a Dual-Path Speech Encoder (DPSE) to disentangle expression-related and head-pose-related features, while an autoregressive Keyframe Establishment Learning (KEL) module predicts the most salient motion frames. These components are integrated into a Dual-path Motion generator to synthesize coherent and realistic facial motions. Extensive experiments on HDTF and VoxCeleb demonstrate that KSDiff achieves state-of-the-art performance, with improvements in both lip synchronization accuracy and head-pose naturalness. Our results highlight the effectiveness of combining speech disentanglement with keyframe-aware diffusion for talking-head generation. The project page is available at: \url{https://kincin.github.io/KSDiff/}.

\end{abstract}
\section{Introduction}
\label{sec:intro}
Audio-driven facial animation has attracted increasing attention multimedia due to its wide applications in digital entertainment, virtual avatars, and human–computer interaction. Recently, diffusion models have demonstrated remarkable capability in synthesizing realistic and temporally coherent talking faces. Despite these advances, most existing methods \cite{stan2023facediffuser, shen2023difftalk, ma2024diffspeaker, wang2024emotivetalkexpressivetalkinghead} treat speech features as a monolithic representation and overlook their fine-grained roles in driving different facial motions. 

\begin{figure}[t]
    \centering
    \includegraphics[width=\linewidth]{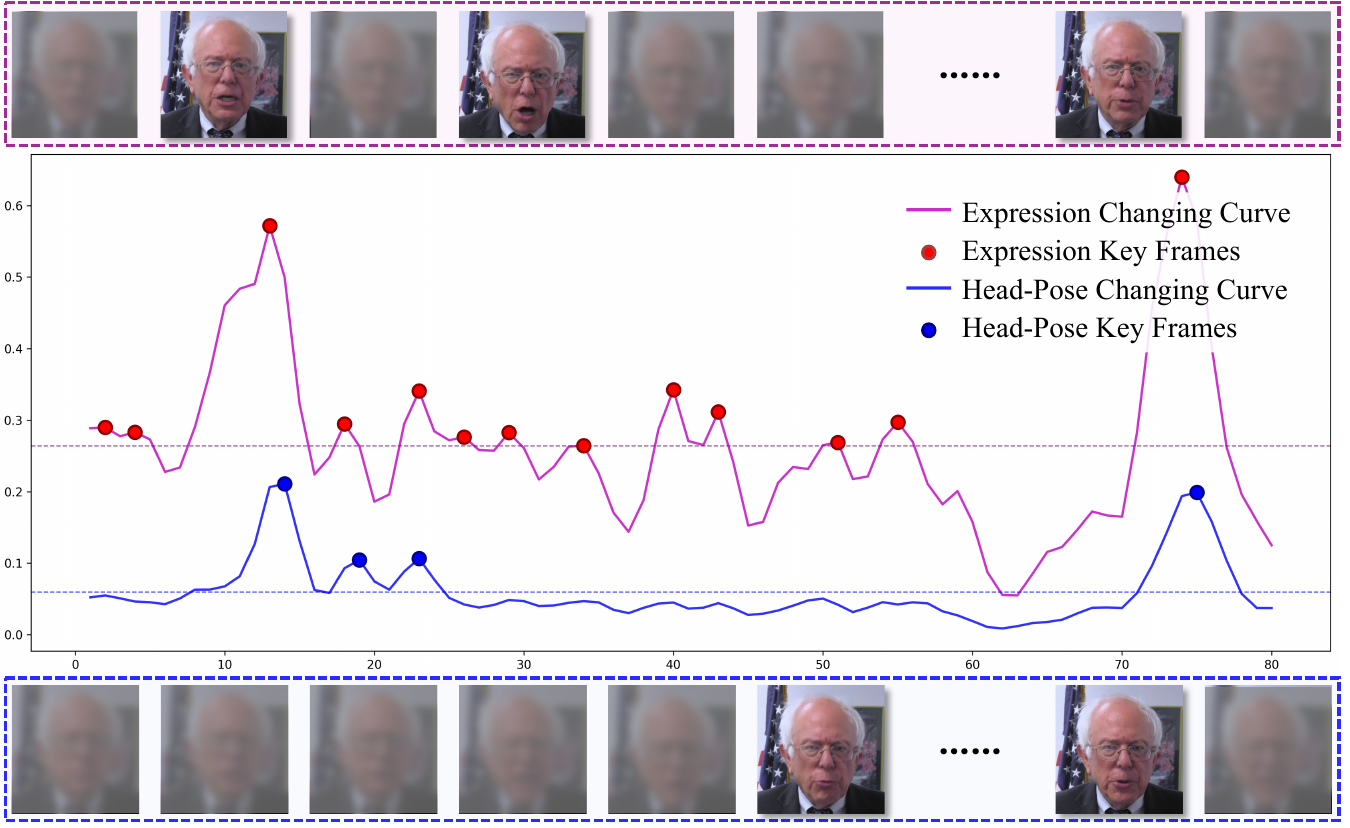}
    \caption{\textbf{Motivation illustration.} Expression features tend to be more dynamic than head-pose features. Specifically, expressions are associated with high-frequency variations, while head poses primarily reflect low-frequency information.}
    \label{fig1}
\end{figure} 
Several recent studies have attempted to address these challenges. SadTalker~\cite{zhang2023sadtalker} and Synctalk \cite{peng2024synctalk} focuses on head-pose modeling but lacks refined speech guidance. ProsodyTalker~\cite{li2025prosodytalker} introduces prosody into generation to explore the role of speech information, but remains unstable under extreme conditions. KeyFace~\cite{bigata2025keyface} emphasizes keyframe selection, yet the chosen frames often lack clear physical meaning. These observations highlight the necessity of disentangling speech representations and designing principled keyframe modeling strategies. 

\begin{figure*}[t]
    \centering
    \includegraphics[width=\linewidth]{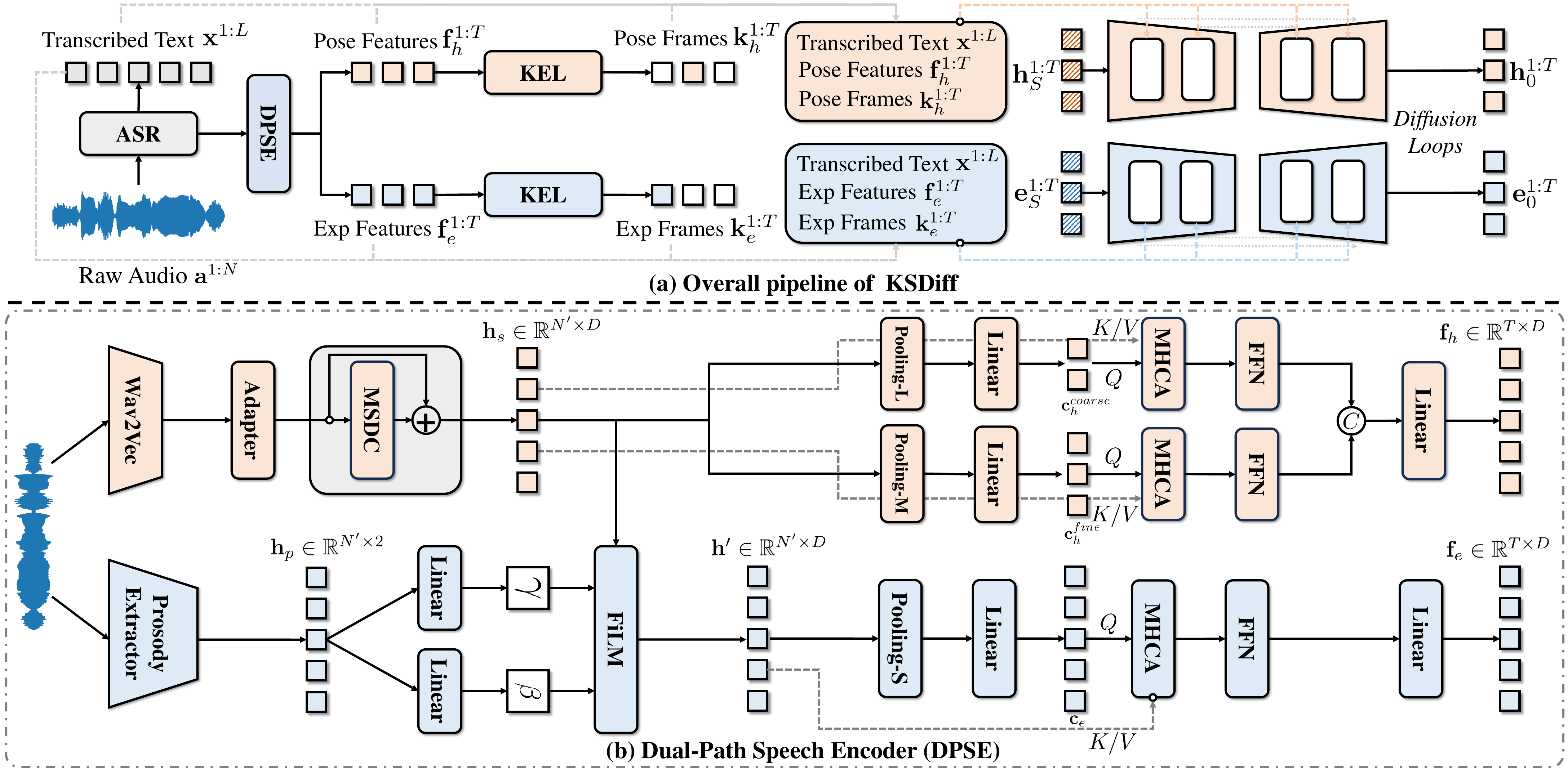}
    \caption{\textbf{Overview of KSDiff.} Input audio is processed by the \textbf{D}ual-\textbf{P}ath \textbf{S}peech \textbf{E}ncoder (DPSE) to disentangle expression- and head-pose-related features. The \textbf{K}eyframe \textbf{E}stablishment \textbf{L}earning (KEL) module extracts corresponding keyframe sequences, which together with the disentangled features are fed into Dual-Path Motion Generators to produce head-pose and expression coefficients.}
    \label{fig2}
\end{figure*}

With the above motivations, we further observe an interesting phenomenon, as illustrated in Fig.~\ref{fig1}: speech features related to expressions typically correspond to high-frequency variations, while head-pose-related features mainly capture low-frequency components. \cite{10887862} Inspired by the concept of anchor points in KeyFace~\cite{bigata2025keyface} and the motion–appearance disentanglement strategy in FD2Talk~\cite{yao2024fd2talk}, we extend this idea by explicitly leveraging keyframe information during generation. To this end, we propose a novel pipeline, KSDiff, a \textbf{K}eyframe-Augmented \textbf{S}peech-Aware Dual-Path \textbf{Diff}usion model for facial animation. Specifically, we design a \textbf{D}ual-\textbf{P}ath \textbf{S}peech \textbf{E}ncoder (DPSE) to disentangle the input audio waveform into expression-related and head-pose-related speech features. These are then processed by the \textbf{K}eyframe \textbf{E}stablishment \textbf{L}earning (KEL) module to generate two corresponding keyframe sequences. Finally, all extracted conditions are integrated into a DiffSpeaker \cite{ma2024diffspeaker} based Dual-Path Motion Generator to synthesize coherent facial motion that jointly captures both expressions and head-pose. Extensive experiments demonstrate that KSDiff achieves state-of-the-art performance across multiple benchmarks.

The main contributions are summarized as follows:
\begin{itemize}[noitemsep, topsep=0pt, leftmargin=*]
    \item We propose a Dual-Path Speech Encoder (DPSE) that disentangles speech features into expression-related and head-pose-related components, facilitating the synthesis of each motion type with precise feature representations.
    \item We introduce a Keyframe Establishment Learning (KEL) that ensures frames with the most intense movements are selected, thereby improving the fidelity of talking.
    \item The proposed dual-path diffusion framework produces highly detailed facial animations with realistic motion dynamics, and extensive experiments validate its effectiveness.
\end{itemize}
\section{Methodology}
The overall framework is illustrated in Fig.~\ref{fig2}. 
Given raw audio $\mathbf{a}^{1:N}$ and transcribed text $\mathbf{x}^{1:L}$, 
we first employ a Dual-Path Speech Encoder (DPSE) to disentangle head-pose-related features $\mathbf{f}_h^{1:T}$ and expression-related features $\mathbf{f}_e^{1:T}$. 
Together with the transcript $\mathbf{x}^{1:L}$, these features are processed by the Keyframe Establishment Learning (KEL) module, 
which generates keyframe sequences $\mathbf{k}_h^{1:T}$ and $\mathbf{k}_e^{1:T}$. 
Subsequently, all components are passed to the DiffSpeaker-based Dual-Path Motion Generator \cite{ma2024diffspeaker}, 
which predicts head-pose coefficients $\mathbf{h}^{1:T}$ and expression coefficients $\mathbf{e}^{1:T}$. 
Finally, DECA \cite{feng2021learning} renders the complete talking-head motion $\mathbf{m}^{1:T}$.
\subsection{Dual-Path Speech Encoder (DPSE)}
\label{subsec1}
Motivated by the observation that expression-related speech cues correspond to high-frequency variations, 
whereas head-pose-related cues are dominated by low-frequency components \cite{hwang2023discohead}, 
we propose a Dual-Path Speech Encoder (DPSE) that explicitly separates speech into two parts \cite{liu2025nerf}. 
Given a raw waveform $\mathbf{a}^{1:N}$, a frozen Wav2Vec encoder $f_{SE}(\cdot)$ \cite{schneider2019wav2vec} extracts frame-level features, 
which are adapted by a lightweight projection layer. 
To capture temporal structures at multiple scales with minimal overhead, 
we employ a parallel Multi-Scale Dilated Convolution (MSDC) block. 
The MSDC consists of $L$ branches with dilations $\{d_\ell\}_{\ell=1}^L$, 
where each branch contains a depthwise $k$-convolution, GroupNorm, GLU gating, and a pointwise convolution. 
Branch outputs are concatenated, projected by a convolution layer, and added residually to produce $\mathbf{h}_s \in \mathbb{R}^{N' \times D}$. 

The hidden speech features $\mathbf{h}_s$ are split into two branches, 
one targeting expression-related cues and the other focusing on head-pose information. 
For the head-pose branch, we apply two windowed pooling operations with different receptive fields: 
a long window $w_h^c$ and a mid-sized window $w_h^f$, 
yielding multi-resolution speech representations. 
Each sequence is further mapped by linear projections to produce coarse and fine head-pose features, 
denoted as $\mathbf{c}_h^{coarse} \in \mathbb{R}^{N_h^c \times D}$ and 
$\mathbf{c}_h^{fine} \in \mathbb{R}^{N_h^f \times D}$.

In parallel, motivated by findings that prosody strongly correlates with expression features 
\cite{li2025prosodytalker, zhao2025spsinger, zhao25d_interspeech}, 
we directly extract $f_0$ and energy from the raw waveform $\mathbf{a}^{1:N}$, 
with per-utterance normalization for robustness. 
The resulting prosody representation $\mathbf{h}_p \in \mathbb{R}^{N' \times 2}$ 
is passed through two parallel linear layers to generate FiLM \cite{perez2018film} conditioning parameters: 
a scaling factor $\gamma$ and a bias term $\beta$. 
These parameters modulate $\mathbf{h}_s$ through a FiLM operation to obtain the prosody-aware speech features $\mathbf{h}' \in \mathbb{R}^{N' \times D}$, 
after which the modulated features are pooled with a short window $w_e$ and passed through a linear layer, 
yielding the expression-related token sequence 
$\mathbf{c}_e \in \mathbb{R}^{N_e \times D}$. 

Subsequently, the speech features with different receptive fields are refined through multi-head cross-attention (MHCA), enabling dynamic interaction across scales. 
Let $\mathbf{C} = \{\mathbf{c}_h^{\text{coarse}}, \mathbf{c}_h^{\text{fine}}, \mathbf{c}_e\}$ denote the multi-scale speech features. For each branch $\mathbf{X}_i \in \{\mathbf{X}_h^{coarse}, \mathbf{X}_h^{fine}, \mathbf{X}_e\}$, the corresponding queries, keys, and values are defined as:
\begin{equation}
    \mathbf{Q}_i = \mathbf{W}_i^{Q}\mathbf{C}_i, \quad
    \mathbf{K}_i = \mathbf{W}_i^{K}\mathbf{H}_i, \quad
    \mathbf{V}_i = \mathbf{W}_i^{V}\mathbf{H}_i,
\end{equation}
where $\mathbf{H} = \{\mathbf{h}_s, \mathbf{h}', \mathbf{h}'\}$ are the hidden speech features aligned with the respective branches, and $\mathbf{W}_i^{Q}$, $\mathbf{W}_i^{K}$, and $\mathbf{W}_i^{V}$ are the learnable parameter matrices.  
The cross-attention output for each branch is computed as:
\begin{equation}
    \mathbf{o}_i = \mathrm{softmax}\!\left(\frac{\mathbf{Q}_i \mathbf{K}_i^\top}{\sqrt{d}}\right)\mathbf{V}_i.
\end{equation}

The outputs of MHCA are subsequently passed through feed-forward networks (FFN) to further enhance representational capacity. The coarse and fine head-pose features are then concatenated and projected via a linear layer to form the head-pose-related representation $\mathbf{f}_h \in \mathbb{R}^{T \times D}$, while the expression branch produces the expression-related representation $\mathbf{f}_e \in \mathbb{R}^{T \times D}$.

\subsection{Keyframe Establishment Learning (KEL)}
\label{subsec2}
Inspired by KeyFace \cite{bigata2025keyface}, we highlight the importance of keyframes in talking-head generation. 
To identify key motion moments, we measure the inter-frame variation of ground-truth head-pose $\hat{\mathbf{h}} \in \mathbb{R}^{T \times 9}$ and expression parameters $\hat{\mathbf{e}} \in \mathbb{R}^{T \times 50}$.  

The head-pose $\hat{\mathbf{h}}$ is decomposed into pose parameters $\mathbf{p}$ and camera parameters $\mathbf{c}$, such that \(
\hat{\mathbf{h}} = [\mathbf{p}, \mathbf{c}], \quad 
\mathbf{p} \in \mathbb{R}^{T \times 6}, \;
\mathbf{c} \in \mathbb{R}^{T \times 3}\). 
Within $\mathbf{p}$, the first three dimensions correspond to the rotation parameters $\mathbf{r} \in \mathbb{R}^{T \times 3}$. The relative rotation between consecutive frames is defined as $\Delta r_{t} = \mathbf{r}_{t} \mathbf{r}_{t-1}^\top$, from which the angular magnitude $\theta_t$ is extracted. The remaining three dimensions of $\mathbf{p}$ represent neck parameters $\mathbf{n} \in \mathbb{R}^{T \times 3}$, which are concatenated with the camera parameters $\mathbf{c}$ to form the combined sequence $\mathbf{c}' = [\mathbf{n}, \mathbf{c}] \in \mathbb{R}^{T \times 6}$. The inter-frame variation of these parameters is measured using the Euclidean distance $\Delta c'_t = \lVert \mathbf{c}'_t - \mathbf{c}'_{t-1} \rVert_2$.  

For expression coefficients $\hat{\mathbf{e}}$, the variation is directly computed as the inter-frame Euclidean distance. The final overall head-pose and expression variation sequences are then given by:
\begin{equation}
\delta_{h,t} = \theta_t + \Delta c'_t, 
\quad 
\delta_{e,t} = \lVert \hat{\mathbf{e}}_t - \hat{\mathbf{e}}_{t-1} \rVert_2.
\end{equation}
Both sequences $\boldsymbol{\delta}^{h} \in \mathbb{R}^{T}$ and $\boldsymbol{\delta}^{e} \in \mathbb{R}^{T}$ are smoothed using a Gaussian filter, and local maxima above a data-dependent threshold are selected as target head-pose keyframes $\hat{\mathbf{k}}_h \in \{0,1\}^{T}$ and expression keyframes $\hat{\mathbf{k}}_e \in \{0,1\}^{T}$.  

To predict key moments, we employ two Transformer-based predictors that autoregressively generate binary keyframe sequences conditioned on speech embeddings and text features. Given the processed speech features $\mathbf{f}_h$, $\mathbf{f}_e$ and the transcribed text $\mathbf{x}$, the predictors produce head-pose keyframes $\mathbf{k}_h$ and expression keyframes $\mathbf{k}_e$. To mitigate the strong class imbalance between sparse keyframes and abundant non-keyframes, we adopt a weighted binary cross-entropy loss:
\begin{equation}
\mathcal{L}_{\text{BCE}}^h = - \sum_t \left( w_1 \hat{k}_{h,t} \log k_{h,t} + w_0 (1-\hat{k}_{h,t}) \log (1-k_{h,t}) \right),
\end{equation}
where $k_{h,t} \in \{0,1\}$ is the predicted probability that frame $t$ is a head-pose keyframe and $\hat{k}_{h,t} \in \{0,1\}$ is the ground truth. The loss for expression keyframes, $\mathcal{L}_{\text{BCE}}^e$, follows the same formulation. The positive class (keyframe) is assigned a larger weight $w_1 > w_0$, while the negative class (non-keyframe) is assigned $w_0$.

\subsection{Dual-Path Motion Generator}
Based on DiffSpeaker \cite{ma2024diffspeaker}, we design a dual-path framework to separately generate head-pose and expression coefficients, denoted as $\mathbf{h}^{1:T}$ and $\mathbf{e}^{1:T}$. 
In the head-pose path, the diffusion process is conditioned on the transcribed text $\mathbf{x}^{1:L}$, head-pose-related speech features $\mathbf{f}_h^{1:T}$, and the head-pose keyframe sequence $\mathbf{k}_h^{1:T}$. 
In parallel, the expression path employs the same conditioning scheme with expression-related speech features $\mathbf{f}_e^{1:T}$ and keyframe sequence $\mathbf{k}_e^{1:T}$. 
Following the DiffSpeaker formulation, the diffusion loss for each path is defined as:
\begin{equation}
\mathcal{L}_{\text{diff}}^h = \lambda_1 \mathcal{L}_{\text{rec}}^h + \lambda_2 \mathcal{L}_{\text{vel}}^h,
\end{equation}
where $\lambda_1=\lambda_2=1$, and $\mathcal{L}_{\text{diff}}^e$ follows an identical formulation.  

To further enhance motion quality, we incorporate a multi-resolution spectral loss and a dynamics regularization term. 
Specifically, we adapt the MR-STFT loss \cite{song2021improved} to kinematic sequences and apply it independently to each branch:
\begin{equation}
\mathcal{L}_{\mathrm{mr}}^h = \sum_{r\in\mathcal{R}} \big\|\,|S_r(\mathbf{h})| - |S_r(\hat{\mathbf{h}})|\,\big\|_1,
\end{equation}
with $\mathcal{L}_{\mathrm{mr}}^e$ defined analogously.  

Finally, the total loss of the proposed KSDiff framework is expressed as:  
\begin{equation}
\begin{aligned}
\mathcal{L}_h &= \lambda_{\text{mr}}^h \mathcal{L}_{\text{mr}}^h 
              + \lambda_{\text{BCE}}^h \mathcal{L}_{\text{BCE}}^h 
              + \lambda_{\text{diff}}^h \mathcal{L}_{\text{diff}}^h, \\
\mathcal{L}_e &= \lambda_{\text{mr}}^e \mathcal{L}_{\text{mr}}^e 
              + \lambda_{\text{BCE}}^e \mathcal{L}_{\text{BCE}}^e 
              + \lambda_{\text{diff}}^e \mathcal{L}_{\text{diff}}^e,
\end{aligned}
\end{equation}
where $\mathcal{L}_h$ and $\mathcal{L}_e$ denote the losses for the head-pose and expression branches, respectively. We set the loss weights as $\lambda_{\text{mr}}=0.3$, $\lambda_{\text{BCE}}=0.5$, and $\lambda_{\text{diff}}=1$ for both branches.  
\section{Experiments Setups}

\begin{table*}[t]
\centering
\caption{Objective comparison of lip synchronization and head motion on the HDTF~\cite{zhang2021flow} and VoxCeleb~\cite{nagrani2017voxceleb} datasets. Best scores are shown in \textbf{bold} and the second best are \underline{underlined}. For clarity, LVE values are scaled by $10^{-5}$ mm.}
\label{tab1}
\resizebox{\linewidth}{!}{
\begin{tabular}{l ccccc ccccc}
\toprule
\multirow{2}{*}{Method}  & \multicolumn{5}{c}{HDTF dataset}  & \multicolumn{5}{c}{VoxCeleb dataset}  \\
\cmidrule(lr){2-6} \cmidrule(lr){7-11}
& LSE-C$\uparrow$ & LSE-D$\downarrow$ & LVE$\downarrow$ & Diversity$\uparrow$ & Beat Align$\uparrow$ & LSE-C$\uparrow$ & LSE-D$\downarrow$ & LVE$\downarrow$ & Diversity$\uparrow$ & Beat Align$\uparrow$\\
\midrule
SadTalker \cite{zhang2023sadtalker}          & 0.625 & 10.121 & 5.918 & 0.246 & 0.274 & 0.653 & 9.981 & 5.802 & 0.296 & 0.305 \\
FaceDiffuser \cite{stan2023facediffuser}     & 0.594 & 11.156 & 6.226 & - & - & 0.627 & 10.530 & 6.091 & - & - \\
DiffTalk \cite{shen2023difftalk}             & 0.689 & 9.884 & 5.279 & 0.281 & 0.295 & 0.706 & 9.743 & 5.026 & 0.297 & 0.324 \\
Hallo2 \cite{cui2024hallo2}                  & 0.704 & 9.629 & 5.437 & \underline{0.293} & 0.302 & 0.711 & 9.841 & 5.174 & \underline{0.316} & 0.347 \\
KeyFace \cite{bigata2025keyface}             & \textbf{0.717} & \underline{9.541} & 5.095 & 0.274 & \underline{0.331} & \textbf{0.732} & \underline{9.415} & 4.821 & 0.310 & \underline{0.354} \\
DiffSpeaker \cite{ma2024diffspeaker}         & 0.702 & 9.916 & \underline{4.926} & - & - & 0.707 & 9.732 & \underline{4.684} & - & -  \\
\midrule
\textbf{KSDiff (Ours)}                       & \underline{0.708} & \textbf{9.204} & \textbf{4.835} & \textbf{0.318} & \textbf{0.354} & \underline{0.713} & \textbf{9.037} & \textbf{4.327} & \textbf{0.328} & \textbf{0.377} \\
\bottomrule
\end{tabular}}
\end{table*}

\begin{table}[t]
\centering
\vspace{-1.5em}
\caption{Subjective evaluation results on full-face quality, lip synchronization, head motion, and fluency.}
\resizebox{\linewidth}{!}{
\begin{tabular}{lcccc}
\toprule
\textbf{\textit{Methods}} & Full-face$\uparrow$ & Lip sync$\uparrow$ & Head motion$\uparrow$ & Fluency$\uparrow$ \\
\midrule
SadTalker \cite{zhang2023sadtalker}        & 3.77 & 3.64 & 4.06 & 3.62\\
FaceDiffuser \cite{stan2023facediffuser}   & 3.24 & 3.36 & 1.58 & 3.37 \\
DiffTalk \cite{shen2023difftalk}           & 4.06 & 3.91 & 4.27 & 4.16\\
Hallo2 \cite{cui2024hallo2}                & 4.05 & 4.31 & 3.98 & 3.94 \\
KeyFace \cite{bigata2025keyface}           & 4.12 & 4.24 & 4.42 & 4.27\\
DiffSpeaker \cite{ma2024diffspeaker}       & 3.69 & 4.32 & 1.37 & 3.63 \\
\midrule
\textbf{KSDiff (Ours)}   & \textbf{4.22} & \textbf{4.48} & \textbf{4.60} & \textbf{4.45} \\
\bottomrule
\end{tabular}
}
\label{tab2}
\end{table}
\begin{table}[t]
\centering
\vspace{-1em}
\caption{Architecture ablation study on the HDTF dataset~\cite{zhang2021flow}. For clarity, LVE values are scaled by $10^{-5}$ mm.}
\resizebox{\linewidth}{!}{
\begin{tabular}{lccccc}
\toprule
\textbf{\textit{Methods}} & LSE-C$\uparrow$ & LSE-D$\downarrow$ & LVE$\downarrow$ & Diversity$\uparrow$ & Beat Align$\uparrow$ \\
\midrule
w/o speech split           & 0.640 & 9.865 & 5.445 & 0.238 & 0.261 \\
w/o dual-path diff         & 0.652 & 9.629 & 5.172 & 0.270 & 0.316 \\
w/o keyframe               & 0.663 & 9.570 & 5.329 & 0.256 & 0.292 \\
w/o prosody                & 0.683 & 9.481 & 4.818 & 0.296 & 0.331 \\
w/o transcript             & 0.699 & 9.372 & 5.720 & 0.305 & 0.342 \\
wav2vec only               & 0.576 & 10.528 & 5.584 & 0.221 & 0.254 \\
\midrule
\textbf{Ours}    & \textbf{0.708} & \textbf{9.204} & \textbf{4.635} & \textbf{0.318} & \textbf{0.354} \\
\bottomrule
\end{tabular}
}
\label{tab4}
\vspace{-1em}
\end{table}
\subsection{Dataset}
We train and evaluate our model on two benchmarks. The High-Definition Talking Face (HDTF) dataset \cite{zhang2021flow} contains high-quality frontal talking-face clips with diverse expressions, making it a standard benchmark. In contrast, the VoxCeleb dataset \cite{nagrani2017voxceleb} includes large-scale speaker videos collected in unconstrained conditions with significant variations in pose, background, and recording quality, serving to evaluate the generalization of our approach.

\subsection{Evaluation Metrics}
We evaluate our model with metrics covering both lip synchronization and head-pose motion. For lip synchronization, \textit{LVE} (Lip Vertex Error)~\cite{richard2021meshtalk} measures the Euclidean distance between predicted and ground-truth lip vertices. \textit{LSE-D} and \textit{LSE-C}~\cite{Prajwal} come from a pre-trained lip-sync discriminator: LSE-D quantifies audio–video embedding distance, while LSE-C reflects discriminator confidence. For head pose, \textit{Diversity}~\cite{Ruiz_2018_CVPR_Workshops} captures the variance of head-pose trajectories, and \textit{Beat Align}~\cite{siyao2022bailando} measures alignment between motion beats and speech accents. Together, these metrics comprehensively assess lip accuracy and head dynamics.

\subsection{Implementation Details}
For both datasets, we preprocess videos with DECA~\cite{feng2021learning} to obtain per-frame expression and head-pose coefficients $\hat{\mathbf{e}} \in \mathbb{R}^{T \times 50}$ and $\hat{\mathbf{h}} \in \mathbb{R}^{T \times 9}$. 
In parallel, \texttt{ffmpeg} is used to extract raw audio, Whisper~\cite{radford2023robust} provides transcribed text $\mathbf{x}$, and prosody features $\mathbf{h}_p$ are obtained as described in Sec.~\ref{subsec1}. 
Keyframe sequences are extracted according to the method in Sec.~\ref{subsec2}. 
All videos are aligned to faces, and audio is resampled to 16kHz.

In the DPSE module, we set the hidden dimension $D=512$, kernel size $k=5$, and apply MSDC followed by a dropout rate of 0.1. The stride $s_n$ is chosen from $\{2,4\}$, and the fused feature dimension is $d_c=512$. We use $w_h^c=1.0$, $w_h^f=0.25$, and $w_e=0.1$. We use the AdamW optimizer for 100k iterations, with 5k warmup steps, a batch size of 32, and a learning rate of $1\text{e}{-4}$. The hidden feature dimension is set to $512$, and the transformer decoder in the keyframe branch has 6 layers and 8 attention heads. The overall training on four NVIDIA RTX A5000 GPUs takes about 16 hours.

\section{Experiments results}
As shown in Table~\ref{tab1}, we compare our KSDiff with other state-of-the-art methods in two categories on HDTF dataset \cite{zhang2021flow} and VoxCeleb dataset \cite{nagrani2017voxceleb}. We use DiffSpeaker \cite{ma2024diffspeaker} as our baseline model, which adopts a diffusion-based Transformer architecture with biased conditional self- and cross- attention mechanisms for speech-driven 3D facial animation. The results highlight the strong capability of KSDiff in capturing fine-grained expression and head-pose details.

To assess perceptual quality, we conduct a user study on four aspects: \textit{1) Full-face naturalness, 2) Lip-sync accuracy, 3) Head motion plausibility, and 4) Overall fluency}. We randomly sample generated videos from different methods and ask 26 participants to rate each aspect on a 5-point Likert scale (1 = very poor, 5 = excellent). 
As shown in Table \ref{tab2}, our KSDiff achieves the highest average scores across all criteria, confirming its superior perceptual quality.

We conduct ablation studies under six settings: 
\textit{1) w/o speech split:} use the entire speech feature for both branches without disentangling expression-related and head-pose–related components; 
\textit{2) w/o dual-path diffusion:} use a single diffusion process for all conditions, directly fusing expression and head-pose information; 
\textit{3) w/o keyframe:} omit the keyframe extraction module; 
\textit{4) w/o prosody:} without prosody guidance in the DPSE module; 
\textit{5) w/o transcript:} without transcript guidance in the pipeline; 
\textit{6) Wav2Vec only:} use raw Wav2Vec features without any refinement modules. 
As shown in Table~\ref{tab4}, the results indicate that each component contributes significantly to the overall performance.

\begin{figure}[t]
    \centering
    \vspace{-0.5em}
    \includegraphics[width=\linewidth]{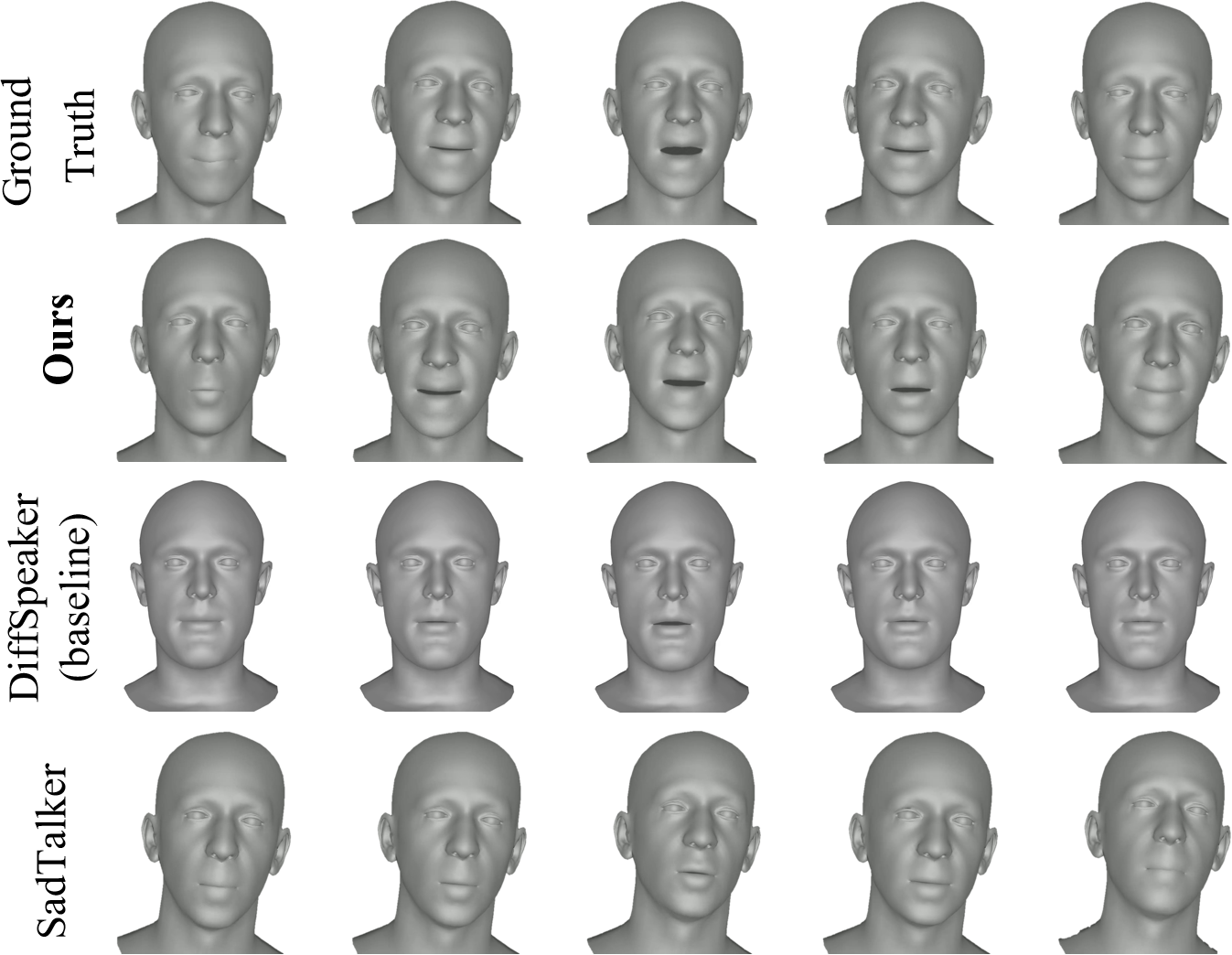}
    \caption{\textbf{Visualization comparison with DiffSpeaker (baseline)~\cite{ma2024diffspeaker} and SadTalker~\cite{zhang2023sadtalker}}. All speakers are uttering the word \textit{``bread''}. Compared to prior methods, our KSDiff generates more natural expressions and richer head-pose dynamics, closely matching the ground-truth sequence. }
    \label{fig3}
    \vspace{-0.2em}
\end{figure} 
As shown in Fig.~\ref{fig3}, KSDiff achieves more accurate head-motion trajectories and natural expressions compared with prior methods. In particular, at key phoneme frames, our model generates plausible head rotations rather than the exaggerated dynamics observed in SadTalker \cite{zhang2023sadtalker}. This leads to more faithful expression–pose coordination and overall results that better align with the ground truth.

\section{Conclusion}
In this paper, we propose KSDiff, a keyframe-augmented speech-aware dual-path diffusion framework for audio-driven facial animation. By disentangling speech into expression- and pose-related features and introducing an autoregressive keyframe learning module, our approach produces natural and coherent facial motions. Experiments on HDTF and VoxCeleb demonstrate that KSDiff achieves state-of-the-art performance in both objective metrics and perceptual quality, while ablation studies validate the contribution of each component. These results highlight the effectiveness and versatility of KSDiff in advancing audio-driven facial animation.

\newpage
\bibliographystyle{IEEEbib}
{\ninept        
\bibliography{KSDiff}
}

\end{document}